\documentclass[twocolumn,prl,superscriptaddress,showpacs,preprintnumbers,floatfix]{revtex4}
\usepackage{amsmath,dcolumn,graphicx}

\begin{document}

\title{A universal relationship between magnetization and local structure
changes below the ferromagnetic transition in La$_{1-x}$Ca$_x$MnO$_3$;
evidence for magnetic dimers.}

\author{L. Downward} \affiliation{Physics Department, University of California, 
Santa Cruz, California 95064, USA}
\author{F. Bridges} \affiliation{Physics Department, University of California, 
Santa Cruz, California 95064, USA}
\author{S. Bushart} \affiliation{Physics Department, University of California, 
Santa Cruz, California 95064, USA}
\author{J.~J. Neumeier} \affiliation{Department of Physics, Montana State
University, Bozeman, MT 59717, USA}
\author{N. Dilley} \affiliation{Quantum Design Inc., 11578 Sorrento Valley
Road, San Diego, CA 92121-1311, USA}
\author{L. Zhou} \affiliation{Stanford Synchrotron Radiation Laboratory,
Stanford Linear Accelerator Center, Menlo Park, CA 94025, USA}

\date{\today}

\begin{abstract}

We present extensive X-ray Absorption Fine Structure (XAFS) measurements on
La$_{1-x}$Ca$_x$MnO$_3$ as a function of B-field  (to 11T) and Ca
concentration, $x$ (21-45\%).  These results reveal local structure changes
(associated with polaron formation) that depend only on the magnetization for
a given sample, irrespective of whether the magnetization is achieved through
a decrease in temperature or an applied magnetic field.  Furthermore, the
relationship between local structure and magnetization depends on the hole
doping.  A model is proposed in which a filamentary magnetization initially
develops via the aggregation of pairs of Mn atoms involving a hole and an
electron site. These pairs have little distortion and it is likely that 
they pre-form at temperatures above T$_c$.

\end{abstract}



\maketitle

In recent years, the interest in manganites has grown significantly.  They
belong to a broader class of materials where charge-spin-lattice interactions
play a crucial role in the observed properties.  An understanding of these
interactions in the manganites may provide insight into other strongly
correlated electron systems, such as the high-T$_c$ superconductors.

The quasicubic manganites (La$_{1-x}$Ca$_x$MnO$_3$) exhibit colossal
magnetoresistance (CMR) for Ca concentrations $x$ $\sim$ 0.2-0.5. Within this
range, there are significant structural changes as a function of temperature
observed at the local atomic level (changes in the Mn-O pair distribution
function (PDF)) and macroscopically (magnetostriction).  The local
distortions are associated with polarons and decrease rapidly through and
below the ferromagnetic transition as observed using
XAFS \cite{XAFS,Booth98b,Lanzara98,Cao00} and
neutron PDF analysis \cite{PDF} at B = 0T.  A tiny B-induced effect at 1T has
also been observed \cite{Field}, but the effect is too small at 1T to explore
the field dependence.

Many have argued for phase separation \cite{Phase_Sep} or two
fluid \cite{Sarker96,Booth98b} models formed of conducting, ferromagnetic
clusters/regions interspersed with poorly conducting regions. Recently,
Ramakrishnan et. al. \cite{Ramakrishnan04} have proposed a two band model of
\emph{coexisting} localized Jahn-Teller (JT) polaron and broad band states.
The idea of phase separation in these materials is generally accepted;
however, the microscopic details of how nanoscale clusters develop into a
fully magnetized state is still poorly understood.  Here we propose a
mechanism in which Mn pairs form filamentary clusters which are interspersed
with Jahn-Teller-distorted, non-magnetic regions.  This is consistent with
Kumar et. al. \cite{Kumar98}, who have shown evidence for small clusters above
T$_c$, where the magnetic susceptibility 
fits a Curie-Weiss law with a Curie constant nearly twice the expected value,
suggesting Mn dimers.  In addition, recent neutron scattering data
\cite{Lynn_priv} show a glass-like phase with short range order above T$_c$,
which is also consistent with the Mn-dimer model presented here.


Previously, we have shown for a few samples \cite{Booth98b,Downward03} that
the decrease in average local distortion, $\Delta (\sigma^2)$, as T is
decreased below T$_c$, is a simple function of the sample magnetization, M.
Here $\sigma$ is the width of the Mn-O PDF obtained from EXAFS, and $\Delta
(\sigma^2)$ is the T- or B-induced change in $\sigma^2$. By extending those
measurements to high fields (9-11T) for some samples, and to a range of Ca
concentrations, $0.21 \le x \le 0.45$, we find that the plots of
$\Delta(\sigma^2)$ versus M (each normalized to their value at low T) depend
primarily on hole concentration. 
 
We find three important results \emph{for each sample}: 
(1) little distortion is removed 
until the fraction of magnetized sites (M/M$_0$) reaches $\sim 2y$, where $y$
is the hole concentration 
(2) above $\sim 2y$, distortions are removed more rapidly as the sample
becomes fully magnetized, and (3) the point of change between these two
behaviors occurs at $M/M_0$ $\sim 2y$.
In addition, for the samples measured in an applied B-Field, the above three
results hold regardless of field, indicating that the change in local structure
depends only on magnetization.

\begin{figure}
\vbox{
\hspace*{-0.05in}
\includegraphics[width=2.5in,height=2.0in,clip]{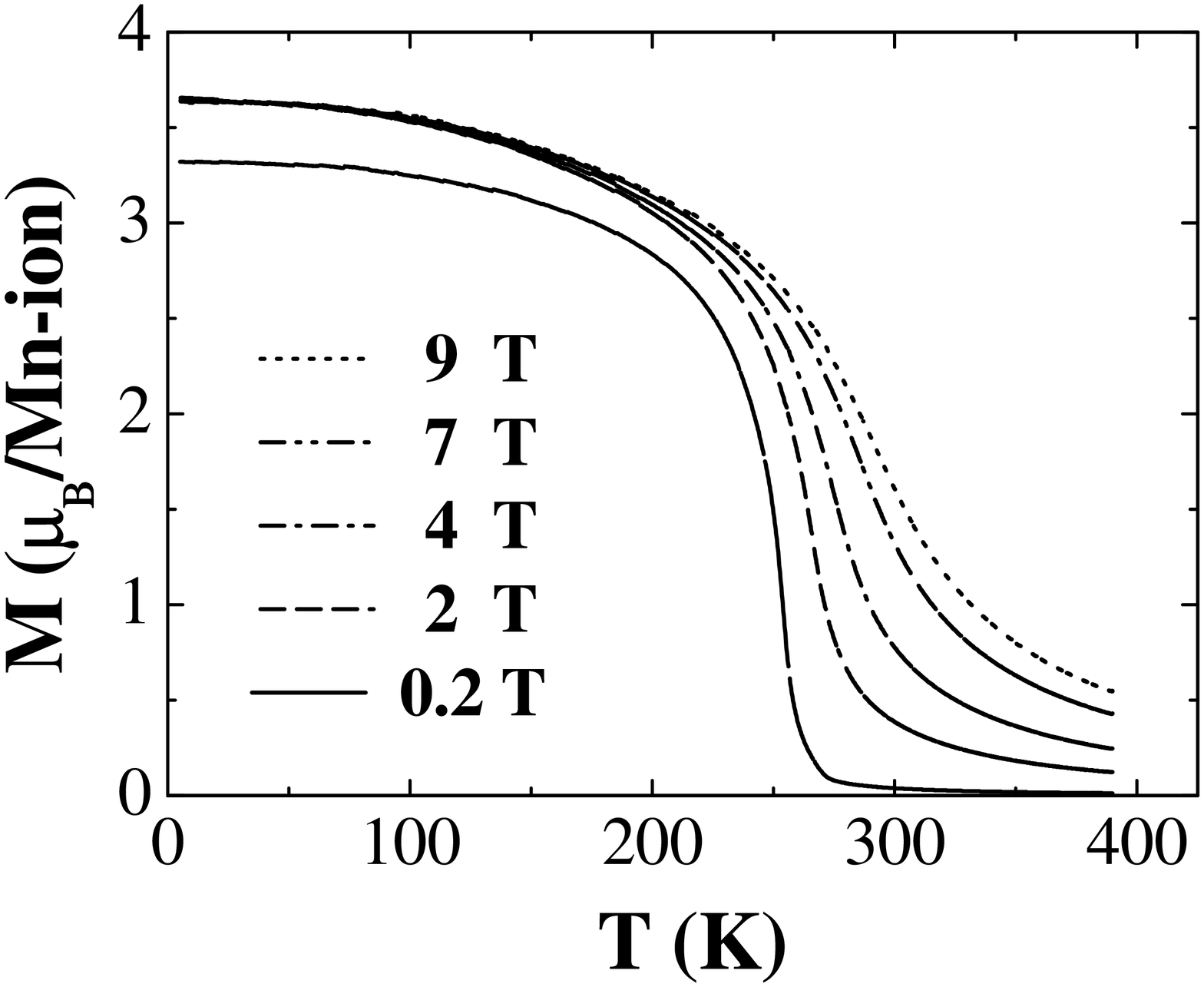}
\includegraphics[width=2.5in,height=2.0in,clip]{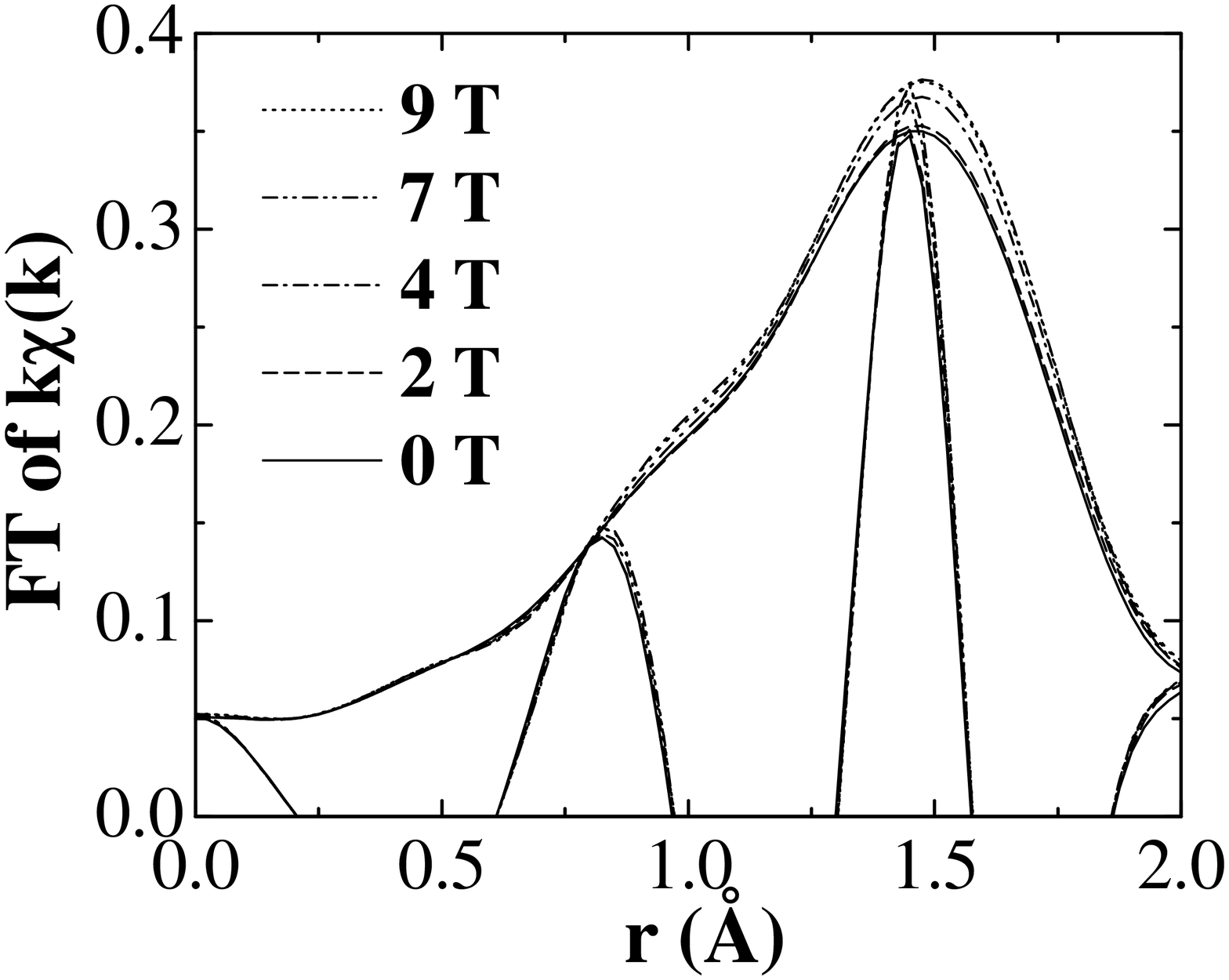}
}

\caption{ Top: Magnetization versus T for La$_{0.7}$Ca$_{0.3}$MnO$_3$.  Bottom: The
r-space peak for the Mn-O bond (La$_{0.7}$Ca$_{0.3}$MnO$_3$) near T$_c$ $\sim$ 260K.
The fast oscillation is the real part of the transform (FT$_R$) while the
envelope is $\pm \protect\sqrt{FT_R^2 + FT_I^2}$, where FT$_I$ is the imaginary
part. 
The FT range is 3.5-11.5\AA, with 0.3~\AA$^{-1}$ Gaussian broadening. } 

\vspace*{-0.1in}
\label{rs}
\end{figure}

\begin{figure}[b]
\includegraphics[width=2.5in,height=2.8in,clip]{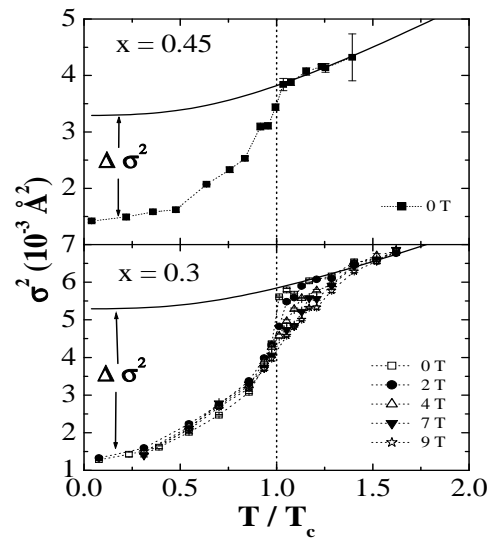}
\vspace*{-0.1in}

\caption{$\sigma ^2$ versus $T/T_c$ for the first Mn-O peak of
La$_{0.55}$Ca$_{0.45}$MnO$_3$ (T$_c$ $\sim$ 250K) and La$_{0.7}$Ca$_{0.3}$MnO$_3$ (T$_c$
$\sim$ 260K).  The dotted lines are guides to the eye.  The solid line is the
behavior if no polaron distortions are removed - i.e.  $\sigma^2$ for CaMnO$_3$
plus a large static distortion.  $\Delta (\sigma ^2$) is defined as the
difference between this line and the experimental data.}

\vspace*{-0.1in}
\label{sigma}
\end{figure}

Transmission XAFS Mn K-edge data were collected on powdered samples at the
Stanford Synchrotron Radiation Laboratory (SSRL) on beam lines 7-2, 10-2 and
6-2, using a Si(111) monochromator.  
The sample preparation has been described earlier \cite{Booth98b} and accurate
hole concentrations, $y$, were determined by iodometric titration with an
uncertainty of $\pm$ 0.005.  For the XAFS samples, the pressed pellets were
reground, passed through a 400-mesh sieve, and then brushed onto scotch tape
which preferentially holds the smaller grains ($\leq$ 5 $\mu$m) in a thin
layer.  Layers of tape were stacked to obtain a Mn K-edge absorption 
step height ($\mu_\mathrm{Mn}t) \sim 0.5$ for each sample.  Generally, four scans
were collected at each T (and B-field) for each sample.  For the B-field
measurements, the field was aligned parallel to the tape layers such that the
demagnetization factor is negligible.  The 
magnetization data were collected on long rods with the B-field parallel to the
axis; the demagnetization factor for this orientation is also negligible.  The
upper panel of Fig. \ref{rs} displays the magnetization at various magnetic
fields for the 30\% Ca sample; an applied B-field broadens the curve and 
shifts it to higher temperature.

In EXAFS analysis, the absorption $\mu_p$ from other atoms (pre-edge
absorption) is removed first, then a sum of splines is used to obtain the
background absorption $\mu_0$ above the edge \cite{Li95b,Boothweb}.  Next the
XAFS function, $\chi$, is obtained as a function of photoelectron wave vector
from $\chi(k) = \mu/\mu_0 - 1$ (where $k = \sqrt{2 m_e (E-E_0)/\hbar^2}$).  See
Fig.  2 of ref.  \cite{Booth98b} for an example of the data quality.  The
$k\chi(k)$ data are then Fourier transformed (FT) to r-space.  An expanded view
of the first Mn-O r-space peak for La$_{0.7}$Ca$_{0.3}$MnO$_3$ for T near T$_c$
at several fields is plotted in Fig. \ref{rs}: bottom.  As B increases, the
amplitude of the Mn-O peak increases; thus the average value of $\sigma$ must
decrease, as observed (see Fig.  \ref{sigma}: bottom). In contrast, for T $\ll$
T$_c$ or T $\gg$ T$_c$, no significant B-field induced change in $\sigma$ is
observed.  Similar B-field induced changes were observed for the 21\% sample
for T near T$_c$.  Changes in the Mn-O peak as a function of temperature also
correlate with T$_c$ for all samples.

The r-space data were fit \cite{Boothweb} to similarly transformed standard
functions, calculated using the \texttt{FEFF6} code \cite{FEFF6}.  We used
E$_0$ and S$^2_0$ determined in earlier studies \cite{Booth98b}, constrained
the number of neighbors to 6, and fit the first Mn-O peak using an average bond
length and a single broadening parameter $\sigma$ \cite{TwoPeakFit}.
This method of fitting provides a single-parameter-measure of the local
disorder for comparisons with M.  For each temperature, separate fits were made
to each of four traces and the average value of $\sigma$ calculated; the rms
fluctuation about the average, gives the relative errors, which are comparable
to the symbol size in most cases (See Fig. \ref{sigma}). The absolute error for
$\sigma^2$ depends on the errors in S$_0^2$ and in the FEFF calculation (both
systematic errors), and may be of order 10-15\%.  This error primarily changes
the static component of $\sigma^2$, and shifts the $\sigma^2$ versus T plot
vertically.

\begin{figure}
\includegraphics[width=2.8in,height=3.3in,clip]{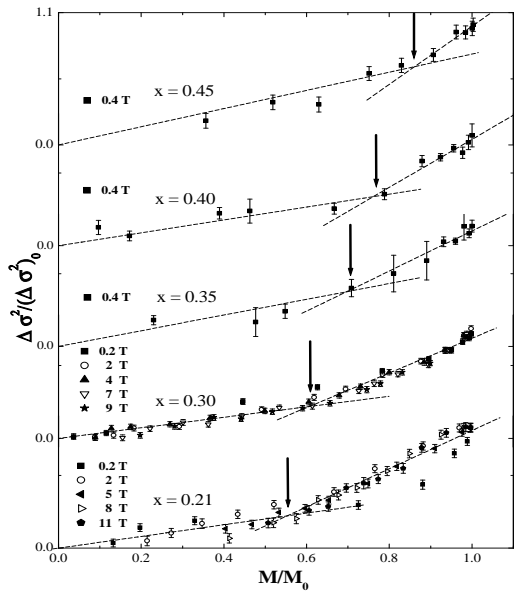}

\caption{$\Delta (\sigma ^2$) versus relative magnetization for the five
different concentrations.  $\Delta (\sigma^2$) and $M$ have been normalized
to their respective values at low T.  $\Delta (\sigma ^2$), defined
in Fig \ref{sigma}, is the decrease in $\sigma ^2$ as T is lowered below
T$_c$ that is attributed to the loss of polaronic distortion.  Note that the
slope of $\Delta (\sigma ^2$) versus $M/M_0$ changes at roughly $M/M_0 \sim 2y$
(see Fig. \ref{bpconc}). }

\vspace*{-0.1in}
\label{Dsigma}
\end{figure}

\begin{figure}[b]
\includegraphics[width=2.25in,height=1.75in,clip]{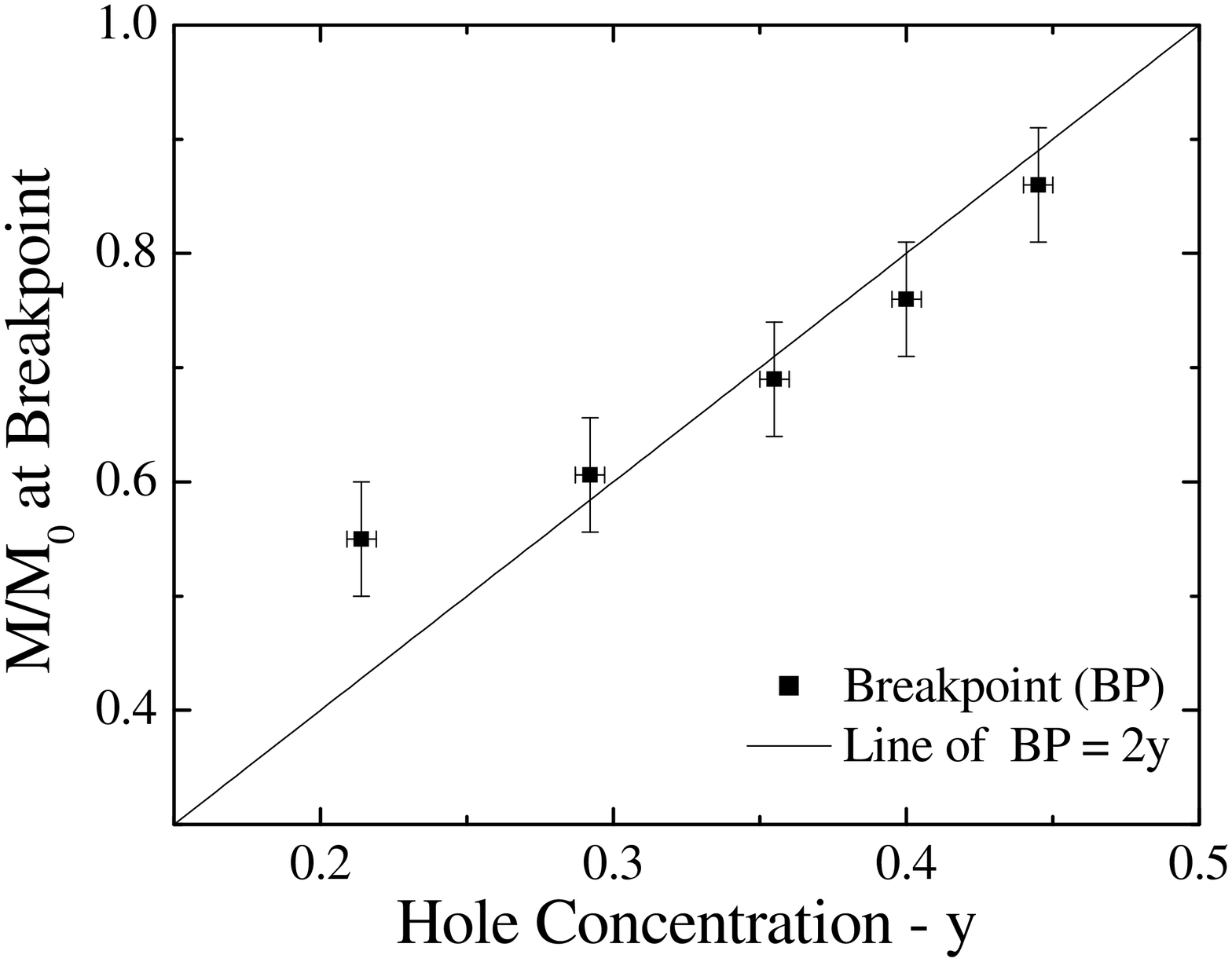}

\caption{Breakpoint in Fig. \ref{Dsigma} versus hole concentration,
$y$.  The breakpoint occurs at $\sim 2y$ for each sample. }

\vspace*{-0.1in}
\label{bpconc}
\end{figure}

$\sigma^2$(T/T$_c$) for the Mn-O peak is plotted in Fig. \ref{sigma} for 
30\%  and 45\% Ca.  For the 30\% sample, the curves shift to
higher T and broaden as B increases, as is also observed in the
magnetization measurements (see Fig. \ref{rs}: top).  

The contributions to $\sigma^2$(T) add up in quadrature when different
broadening mechanisms are uncorrelated, i.e. $\sigma^2(T) =
\sigma^2_{phonons}(T) + \sigma^2_{static} + \sigma^2_{polaron}(T)$,
%
%
where $\sigma^2_{phonons}$(T) is the contribution from thermal phonons,
$\sigma^2_{static}$ arises from temperature independent static distortions, and
$\sigma^2_{polaron}$(T) is the contribution from the presence of hopping
polarons, at 300K (which we will argue below, are likely hopping ``dimerons").
The latter becomes T-independent when M $\sim$ 0 - we refer to this as
$\sigma^2_{RT-polaron}$; the remaining weak T dependence just above T$_c$ is
nearly identical to that for pure CaMnO$_3$ \cite{Booth98b} and arises from the
thermal phonon contributions; the solid line in Fig. \ref{sigma} is the sum of
the phonon and RT-polaron distortion.  The difference between this line and the
data is $\Delta (\sigma^2$(T)) and represents the amount of polaron distortion
removed (Fig.  \ref{sigma}).  Using $\Delta (\sigma^2$(T)) and M(T), we can
plot the changes in the local distortions as a function of M.
In Fig.  \ref{Dsigma}, we show $\Delta(\sigma^2$) versus M, with each
normalized to their low-T values.  The data follow the same linear
relationship, within the errors, irrespective of whether the magnetization was
obtained by a change in T or B (for x = 0.3 and 0.21).  Thus, for a given
sample, the distortion removed is a nearly universal function of M.  An
important feature of every plot is the low initial slope; little distortion is
removed until the sample is more than 50\% magnetized.  Then $\Delta
(\sigma^2$(T)) increases rapidly with M - i.e. M and $\Delta (\sigma^2$) are
not quite in phase.  The breakpoint occurs at M/M$_0$ $\sim 2y$ for each sample
(see Fig.  \ref{bpconc}) \cite{LCMO_21},
i.e.  twice the number of holes. The breakpoint at $\sim 2y$ and the low slope
for M/M$_0$ $< 2y$ suggests that the magnetization develops via pairs of Mn
sites - a hole and a (distorted) electron site. We propose that such dimer
pairs (which we call a dimeron) form at some temperature T$^*$
\cite{Patanjali99} above T$_c$, which likely corresponds to the temperature at
which there is a break in the susceptibility plots. \cite{Kumar98}

\begin{figure}
\vbox{
\includegraphics[width=2.0in,clip]{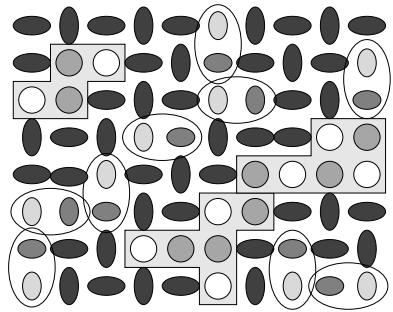}
}

\caption{Model of magnetization process for T $\sim$ T$_c$ (only Mn sites
shown).  Black ovals - JT-distorted polarons; encircled pairs - dimerons
consisting of a partially distorted electron/hole pair; outlined regions -
magnetic clusters with magnetized hole sites (open circles) and magnetized electron sites, with JT-distortion removed (gray circles).}

\vspace*{-0.1in}
\label{model}
\end{figure}

In this dimeron model, an electron must be able to rapidly hop back and forth
between two sites for the double exchange mechanism to be operative;
this is only feasible for an electron/hole pair.
For the Jahn-Teller (JT) distorted site it costs energy
E$_{JT}$ to completely undistort the site; consequently, as long as holes are
present, the least amount of energy is required when the magnetization develops
in electron/hole pairs. The dimeron quasi-particle
will also be mobile; the hole changes electron partners as it moves
through the sample.

If we let the distortion per JT-site be a constant,
$\alpha$, then the maximum average distortion removed per magnetized Mn site
(for decreasing T or increasing B-field) is only $\alpha$/2 
in the initial stages of magnetization.  Once the holes are used up, the
distortion removed per Mn site would then increase to $\alpha$.  This is
qualitatively close to the behavior of the data shown in Fig.  \ref{Dsigma};
however, the ratio of the slopes is closer to 1:4, instead of 1:2, as would be
expected from this simple calculation.
However for the dimeron, the electron is partially delocalized and the JT-like
distortion will be reduced, although not eliminated.
As a result, the ratio of the slopes would be larger than 1:2.  For example, if
the dimeron contains half the total polaron distortion of an electron/hole
pair, then the distortion per site would be $\alpha/4$ and the ratio of slopes
would be 1:4.  

Another explanation for the large change in slope could be a variation in the
size of the polaron distortions (different values of E$_{JT}$) throughout the
system, due to strains and local variations in Ca content.  The sites with the
smallest distortion would then magnetize first.  Since EXAFS gives a measure of
the average local distortion over all sites, we cannot easily distinguish
between these two possible explanations, unless at higher temperatures the
dimerons dissociate to form isolated Mn moments \cite{Kumar98} with larger J-T
polaron distortions.  Further experiments are underway to investigate this
possibility. 
Also, as T approaches T$_c$ it is possible that small clusters of
dimerons can form - the EXAFS data are not in conflict with such a possibility,
they only indicate that equal numbers of electrons and holes (i.e. an integer
number of pairs) have aggregated in the magnetic clusters. The EXAFS data,
however, are not consistent with a large number of Mn triplets (e.g. two
electrons and one hole or two holes and one electron). 

A ``cartoon" of this model for T $\sim$ T$_c$ is depicted in Fig.  \ref{model}.
Pairs of partially distorted holes (small light gray ovals) and partially
distorted electron Mn sites (small dark gray ovals) are encircled at some
instant in time, to represent a dimeron; these are continuously hopping as the
hole changes electron partners. Several magnetized clusters are also shown; in
each cluster, the dimerons become completely undistorted and lose their
identity as the electron becomes much more delocalized (initial electron sites
represented by gray circles; initial hole sites by open circles)
\cite{Mag_cluster}. 
At low M (M/M$_0$ $< 2y$), the dimerons in the non-magnetic regions are
continually hopping and will occasionally diffuse close enough to a magnetic
cluster that they can join the cluster.  Here the quenched disorder from the
positions of the Ca constrains how the magnetization develops, because for
charge neutrality, the holes cannot be too far away from Ca sites.
Consequently, if the magnetic clusters for M/M$_0 < 2y $ are formed from
dimeron aggregation, then there will be filaments of the excess JT-distorted
electron sites (black ovals) throughout the magnetized regions.  Thus, the
magnetic cluster development in this regime may be similar to diffusion limited
aggregation, with the majority of the magnetic thermal fluctuations occurring
in the paramagnetic and boundary regions.  It will lead to interpenetrating
nanoscale filamentary clusters of magnetic and non-magnetic regions.  

In summary, XAFS data as a function of B and T indicate that the decrease in
lattice distortion with increasing M is a nearly universal function for each
sample.  The decrease in distortion is small up to M/M$_0$ $\sim 2y$, which
suggests that the magnetic clusters develop by the aggregation of Mn pairs
(electron/hole sites), which we call dimerons. The location of such aggregated
dimerons is constrained by the quenched disorder of the Ca distribution,  which
leads to filamentary magnetic clusters for M/M$_0$ $< 2y$.  Increased local
distortions should be found at higher temperatures when the dimerons
dissociate.

\acknowledgments 

The work at UCSC was supported in part by NSF grant DMR0301971. We thank A.
Millis and P Littlewood for helpful discussions. The experiments were
performed at SSRL, which is operated by the DOE, Division of Chemical
Sciences, and by the NIH, Biomedical Resource Technology Program, Division
of Research Resources.

\vspace*{-0.1in}
\bibliographystyle{prsty}

\bibliography{/home/users/lisa/papers/bib/bibli}

\begin{thebibliography}{10}
\bibitem{TwoPeakFit}
Modeling with two Mn-O peaks (ref. \cite {Lanzara98}) requires two amplitudes
  and two coupled $\sigma $'s, thus, a single parameter is not available for
  comparison to the magnetization.
\bibitem{Mag_cluster}
At this point, the electrons in the magnetic cluster are delocalized, but we
  keep pairs of (undistorted) open and gray circles for counting purposes.
\end{thebibliography}

\end{document}